\documentclass[twocolumn,showpacs,twocolumn,floatfix,superscriptaddress]{revtex4}
\usepackage{times,amsmath,amsfonts,amssymb,latexsym}
\usepackage{graphicx}
\usepackage{dcolumn}
\usepackage{bm}

\newcommand{\ket}[1]{| #1 \rangle}

\begin{document}
\title{Fidelity in topological quantum phases of matter}
\author{Silvano Garnerone}
\email{garneron@usc.edu}
\affiliation{Department of Physics and Astronomy, Center for Quantum Information Science\& Technology, University of Southern California, Los Angeles, CA 90089}
\author{Damian Abasto}
\affiliation{Department of Physics and Astronomy, Center for Quantum Information Science\& Technology, University of Southern California, Los Angeles, CA 90089}
\author{Stephan Haas}
\affiliation{Department of Physics and Astronomy, Center for Quantum Information Science\& Technology, University of Southern California, Los Angeles, CA 90089}
\author{Paolo Zanardi}
\altaffiliation[Also at ]{Institute for Scientific Interchange, Viale Settimio Severo 65, I-10133 Torino, Italy}
\affiliation{Department of Physics and Astronomy, Center for Quantum Information Science\& Technology, University of Southern California, Los Angeles, CA 90089}
\date{\today}
\begin{abstract}

Quantum phase transitions that take place between two distinct topological phases remain an unexplored
area for the applicability of the fidelity approach. Here, we apply this method to spin systems in two and three
dimensions and show that the fidelity susceptibility can be used to 
determine the boundary between different topological
phases particular to these models, while at the same time offering information about the critical exponent of the
correlation length. The success of this approach 
relies on its independence on local order parameters or breaking symmetry
mechanisms, with which non-topological phases are usually characterized. We also consider a
topological insulator/superconducting phase transition in three dimensions and point out the relevant features 
of fidelity susceptibility at the boundary between these phases. 
\end{abstract}
\maketitle
\section{Introduction}
In recent years topological phases have been intensively studied in 
condensed matter systems. These exotic 
states appear in different contexts, such as fractional 
quantum Hall physics, spin liquids and topological insulators \cite{Wen, ReSa, Kit,TKN,KaMe,YaKi,BHZ}. 
Their understanding is relevant
to topological quantum computation, 
which provides the 
paradigm to store and manipulate information in topologically 
non-trivial systems \cite{NaSiSt}. In a seminal work \cite{Kit} Kitaev 
introduced a spin model which can be exactly solved 
using a mapping to Majorana fermions coupled to a static $ \mathbb{Z}_2 $ 
gauge field ( \cite{Pac}, see also Ref. \cite{ViScDu} for a perturbative approach). 
Generalizations of the Kitaev model 
with respect to lattice geometry, spatial dimension and 
local Hilbert space dimension have appeared recently \cite{YaKi,Ryu,YZS}. 
In some regions of parameter space these models 
have non-trivial topological properties which cannot be described by 
any local order parameter. This prevents the direct applicability 
of the Landau-Ginzburg paradigm for the study of their phase transitions. 
Alternative ways of understanding criticality in systems with no local 
order parameter have been suggested in connection to quantum information. 
In particular, entanglement entropy (for a review see Ref.\cite{AmFaOs}) 
and the fidelity approach \cite{ZaPa,Zhou,ZGC,ZGC1,CaZa} have attracted a lot 
of attention (for a review see Ref.\cite{Gu}). The reason why these two quantities can describe topological phases 
or determine their boundaries is due to the fact that they depend only on the 
properties of the ground state of the system and do not require 
a priori knowledge of any order parameter.

In this paper we 
shall focus on the fidelity approach to topological phase 
transitions \cite{HaZhHa,AbZa, AbHaZa}. Much work has been done in understanding the 
nature of the phases in the Kitaev honeycomb model and, in 
particular, its fidelity has been studied for the first time 
in Refs. \cite{YaGuSu,ZhZh1}. The present work aims to provide a 
more general understanding of fidelity in topological phase 
transitions. For this porpuse we consider two- and three-dimensional extensions
of the Kitaev honeycomb model. The models we chose differ 
from Kitaev's model in the geometry of the lattice, 
the interactions or the dimension of the ambient space and of the local 
Hilbert space.

\section{Models and Method}
The dramatic changes in a many-body ground state $\ket{\Psi_0(\lambda)}$ across a quantum phase transition can be
captured by the fidelity $F = |\langle \Psi_0(\lambda)|\Psi_0(\lambda+\delta\lambda)\rangle|$ between two ground states
corresponding to slightly different values of the set of parameters defining the Hamiltonian $H(\{\lambda\})$.
Alternatively, one can use
the fidelity susceptibility $\chi = \lim_{\delta\lambda\rightarrow 0}\frac{-2\ln F}
{\delta\lambda^2}$ \cite{YoLiGu}. 
The fidelity quantifies how different two quantum states are. Given the drastic changes that take
place across a quantum phase transition, one should expect the fidelity for nearby states to exhibit a drop across
the boundary phase, signaling an enhanced distinguishability. Indeed, while in the thermodynamic limit the fidelity
(susceptibility) has a drop (divergence) at the quantum critical point, for finite size systems this behavior is
translated in the fidelity susceptibility being extensive away from 
criticality, while superextensive at criticality when 
the operator driving the transition is sufficiently relevant \cite{CaZa}.
Moreover, a scaling analysis of the fidelity susceptibility can be performed to extract the critical exponent $\nu$
of the correlation length $\xi$.

Here we apply this tool from quantum information theory to study the quantum phase transition between two distinct
topological phases in three models: mosaic models, Kitaev's honeycomb model with an external
magnetic field, and  a 3D model exhibiting a transition between two distinct non-abelian 
topological phases. Let us first review them very briefly. 

\begin{figure}[htp]
\centering
\includegraphics[angle=0,scale=0.35]{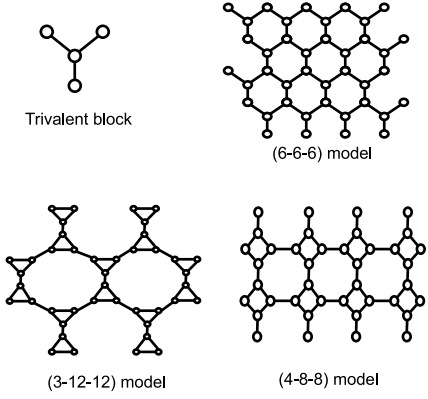}
\caption{The trivalent vertex and the three different mosaic models constructed with it, which are 
considered in the present work. }
\label{fig:fig1}
\end{figure}

\begin{figure}[htp]
\centering
\includegraphics[scale=0.35]{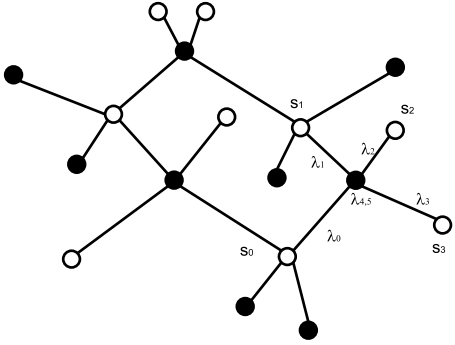}
\caption{The diamond lattice used in the three-dimensional extension 
of the Kitaev model. }
\label{fig:fig2}
\end{figure}

An elegant way to look at the Kitaev model as a special instance 
of a more general class of models is provided by the mosaic classification \cite{YZS}. 
Mosaic models are defined on two-dimensional Bravais lattices constructed with 
a trivalent vertex building block, constrained by translational and rotational symmetry. 
The trivalent vertex is the local border between three polygons (see Fig.\ref{fig:fig1}). The number of 
edges of the three polygons $(e_1,e_2,e_3)$ specifies the mosaic model and 
only four possible such lattices are allowed: $(6-6-6)$, corresponding
to the Kitaev model; $(3-12-12)$, corresponding to the model studied in \cite{YaKi}, 
$(4-6-12)$ and $(4-8-8)$.

The Hamiltonian of all mosaic spin models is given by

\begin{equation}\label{H}
H=-\sum_{u=x,y,z}J_u\sum_{(i,j)\in S(u)} \sigma_i^u\sigma_j^u,
\end{equation}

\noindent where $S(u)$ is the set of edges in the $u$-direction. 
At every site the Hilbert space associated with
the spin-1/2 particle is identified with a two-dimensional physical subspace $\mathcal{M}$ of a four-dimensional Fock
space $\mathcal{\tilde{M}}$, introduced to solve the model. The Pauli operators acting on  $\mathcal{M}$ are
represented by four Majorana operators on $\mathcal{\tilde{M}}$: $\sigma^x=ib^xc$, $\sigma^y=ib^yc$ and
$\sigma^z=ib^zc$, with $\alpha^2=1$, $\alpha\beta = -\beta\alpha$, for $\alpha$, $\beta$ $\in\{b^x,b^y,b^z,c\}$, and
$\alpha\ne\beta$. The physical subspace is obtained through a projection: $\ket{\psi} \in \mathcal{M}
\Longleftrightarrow D\ket{\psi}=\ket{\psi}$, with $D=ib^xb^yb^zc$. The Hamiltonian (\ref{H}) can be rewritten as $H =
\frac{1}{2} \sum_{i,j} \hat{A}_{i,j}c_i c_j$, with the operators $\hat{A}_{ij}=iJ_u\hat{Z}_{ij}$, where
$\hat{Z}_{ij}=ib_i^ub_j^u$ and  
$(i,j)\in S(u)$ ($\hat{A}_{ij}=0$ otherwise). The operators $\hat{Z}_{ij}$ commute with each other and with the
Hamiltonian, so the space of the multi-spin system $\mathcal{L}$ can be decomposed as a direct sum $\mathcal{L} =
\bigoplus_Z \mathcal{L}_Z$, where each sector is indexed by a set of eigenvalues $\{z_{ij}=\pm 1 |(i,j)\in S(u), u
=x,y,z\}$ of the operator $\hat{Z}_{ij}$. Within each sector, the Hamiltonian reduces to a quadratic form in the
Majorana operators $c_i$. The couplings of the Hamiltonian depend on a choice of vortex configuration, given by
the eigenvalues of the plaquette operators $W_{p(n_i)}=-\prod_{i,j \in\partial p(n_i)}Z_{ij}$, one for each of the
$n_i$ isogons $\left( \partial p\left( n_i \right) \right.$ is the set of links belonging 
to plaquette $\left. p\right(n_i \left) \right)$. 
The operators $W_{p(n_i)}$
commute with each other and with the Hamiltonian and since $W_{p(n_i)}^2=1$, their eigenvalues are $w_p=\pm 1$. 
A plaquette with $w_p=+1$ is a vortex free plaquette, while $w_p=-1$ corresponds to a vortex. 
In the following, we
restrict ourselves to the vortex free sector, with a choice of $z_{ij}$ that preserves translational symmetry. 
The ground state energy is at a minimum for this configuration \cite{Lie}. 

Another modification of the Kitaev model that preserves its 
topological nature is the addition of a three-spin interaction 
of the form

\begin{eqnarray}\label{HP}
H=-\sum_{u=x,y,z}J_u\sum_{(i,j)\in S(u)} \sigma_i^u\sigma_j^u- K\sum_{i,j,k}\sigma^x_i\sigma_j^y\sigma_k^z,
\end{eqnarray}

\noindent where the K-term is obtained through a perturbative expansion of a weak (Zeeman) magnetic field $V =
\mathbf{h\cdot\sigma}$ \cite{LaKeCa}. In this case $K \sim \frac{h_xh_yh_z}{J^2}$ and this
Hamiltonian is assumed to approximate the one with a Zeeman term when $K\ll J$, with $ J_u=J $. The inclusion of this
magnetic
perturbation to the original model generates a topological phase with nonabelian anyonic 
excitations in the regime $
|J_z|\le |J_x| + |J_y|, \; |J_y|\le |J_z| + |J_x|, \; |J_x|\le |J_y|+|J_z|$; while outside this region the excitations remain abelian.

Recently analogues of the Kitaev model on three-dimensional 
lattices have been constructed \cite{Ryu,WuArHu}. In the 
present work we focus on the model introduced in Ref. \cite{Ryu}
since it presents quantum phase transitions 
between distinct non-trivial topological phases that, to our knowledge, 
have never been investigated with the fidelity approach. The model is defined on 
a diamond lattice and has a four-dimensional local Hilbert space (see Fig.\ref{fig:fig2}). 
The Hamiltonian is the following
\begin{equation}\label{eq:H_3d}
H=-\sum_{u =0}^{3} J_{u}\sum_{(i,j)\in S(u)} \sigma_{j}^{u} \sigma_{k}^{u} 
(\tau_{j}^{x}\tau_{k}^{x}+\tau_{j}^{z} \tau_{k}^{z}),
\end{equation}
where $ \sigma $ and $ \tau $ are Pauli spin matrices associated to the two local spin-1/2 
degree of freedoms at each point in the lattice. As shown in \cite{Ryu} Hamiltonian (\ref{eq:H_3d}) 
can be mapped to a free Majorana Hamiltonian which, in its ground-state sector, is given by
\begin{equation}\label{eq:H_3d_Mayora}
H=i \sum_{u =0}^{3} J_{u}\sum_{(i,j)\in S(u)} (\lambda_{j}^{4} \lambda_{k}^{4}+\lambda_{j}^{5}\lambda_{k}^{5}),
\end{equation}
and $ \lambda_i^{4,5} $ are Majorana fermions introduced in the representation of the 
Pauli spin matrices (see Ref. \cite{Ryu} for details), analogously to the previous model Hamiltonians.
The addition of next-nearest-neighbour interactions into (\ref{eq:H_3d_Mayora}) allows for the presence of distinct 
non-trivial topological phases in the phase diagram, as we shall see in the next section.

The spin realization of all the above models can be mapped onto a free Majorana fermion Hamiltonian
\begin{equation}\label{eq:H_majora}
H(A)=\frac{i}{4} \sum_{i,j} A_{i,j} c_i c_j, 
\end{equation}
where $ A_{i,j} $ encodes the lattice structure and the parameters of the model.
Once a unit cell of size $s$ has been chosen the Hamiltonian can be written 
in direct space as
\begin{equation}\label{eq:H_cell}
H(A)=\frac{i}{4} \sum_{n,\nu;m,\mu} A_{n,\nu;m,\mu} c_{n,\nu} c_{m,\mu},
\end{equation}
where $ n$ and $m$ are the positions of the unit cells in the lattice and 
$\mu$ and $\nu$ are the positions of the vertices inside the unit cell.
Due to translational invariance $ A $ depends only on $ \nu,\mu $ and $ m-n $. Using a Fourier 
transform, we obtain
\begin{equation}\label{eq:H_bz}
H=\sum_{\textbf{k}} \boldsymbol{\Psi}^{\dagger}(\textbf{k}) \mathcal{H}(\textbf{k}) \boldsymbol{\Psi}(\textbf{k}),
\end{equation}
with $ \Psi_{\alpha}(\textbf{k})= 1/\sqrt{N}\sum_{\textbf{r}} e^{-i \textbf{k}\cdot \textbf{r}} c_{\textbf{r},\alpha}$ 
a complex fermion, $N = sL^d$ the system size, $L$ the number of unit cells in one direction 
of the d-dimensional lattice and $s$ the size 
of the unit cell, 
with $ {\alpha} \in \{1,2,..,s\} $. 
Then in general we have to deal with a free-fermion model characterized by $s$ 
different bands.
The ground-state is obtained filling 
the Fermi sea, where all the levels with negative energy $ \epsilon_{\beta}(\textbf{k})<0 $ are occupied,
\begin{equation}\label{eq:gs}
 \ket{\Psi_0}  \equiv \prod_{\beta,\textbf{k}} b_{\beta}^{\dagger}(\textbf{k})\ket{0}
\end{equation}
 if $ \epsilon_{\beta}(\textbf{k})<0 $.
$ b_{\beta}^{\dagger}(\textbf{k})\equiv \boldsymbol{\Psi}^{\dagger} (\textbf{k})\cdot \textbf{V}_{\beta}(\textbf{k})$,
where $ \textbf{V}_{\beta}(\textbf{k}) $ is 
the eigenvector associated to the $ \epsilon_{\beta}(\textbf{k}) $ eigenvalue of $ \mathcal{H}(\textbf{k}) $. 
The matrix $ U(\textbf{k}) $ diagonalizing $ \mathcal{H}(\textbf{k}) $ has as column vectors $ V_{\beta}(\textbf{k})$. 
The spectrum of the single-body Hamiltonian $ \mathcal{H}(\textbf{k}) $ is symmetric around zero.

The $\textbf{k}$-component of the many-body groundstate $ \ket{\Psi_0}=\prod_{\textbf{k}}\ket{\Psi_0}_{\textbf{k}} $ 
in first quantization is given by 
a functional Slater determinant. 
The $j$-th component of the eigenstate corresponding to the $\beta$-th band
is represented by $V_{j,\beta}(\textbf{k})$.
Denoting with $ s_n $ the number of negative single particle energy bands, 
the first quantized wave function at fixed $\textbf{k}$ is the Slater determinant 
of $s_n$ particles that can occupy $s_n$ bands
\begin{equation}
\langle j_1,...,j_{s_n}|\Psi_0 \rangle_{\textbf{k}}
=
\frac{1}{\sqrt{s_n!}}
\left|
\begin{array}{ccc}
{V}_{j_1,1}(\textbf{k}) & \cdots & {V}_{j_1,s_n}(\textbf{k}) \\ 
\vdots & \ddots & \vdots \\ 
{V}_{j_{s_n},1}(\textbf{k}) & \cdots & {V}_{j_{s_n},s_n}(\textbf{k})
\end{array} 
\right|,
\end{equation}
with $j_i \in \{1,2,...,s\}$.
For these particular models, the fidelity corresponds to the 
product over all $ \textbf{k} $ of the 
absolute value of the overlap of two
Slater determinants at different parameter values 
$\Vert {\langle \tilde{\Psi}_0 | \Psi_0 \rangle}\Vert=\bigotimes_{\textbf{k}}
\Vert{\langle \tilde{\Psi}_0 | \Psi_0 \rangle}_{\textbf{k}}\Vert$ 
with
\begin{eqnarray} 
{\langle \tilde{\Psi}_0 | \Psi_0 \rangle}_{\textbf{k}}
=
\frac{1}{s_n!}\sum_{j_1,...,j_{s_n}}
\left|
\begin{array}{ccc}
\tilde{{V}}_{j_1,1}(\textbf{k}) & \cdots & \tilde{{V}}_{j_1,s_n}(\textbf{k}) \\ 
\vdots & \ddots & \vdots \\ 
\tilde{{V}}_{j_{s_n},1}(\textbf{k}) & \cdots & \tilde{{V}}_{j_{s_n},s_n}(\textbf{k})
\end{array} 
\right|^{*} \nonumber
\\
\times
\left|
\begin{array}{ccc}
{V}_{j_1,1}(\textbf{k}) & \cdots & {V}_{j_1,s_n}(\textbf{k}) \\ 
\vdots & \ddots & \vdots \\ 
{V}_{j_{s_n},1}(\textbf{k}) & \cdots & {V}_{j_{s_n},s_n}(\textbf{k})
\end{array} 
\right|.
\end{eqnarray}

The previous expression, 
using the properties of determinants, can be 
rewritten as (for a proof see pag.291 of Ref.\cite{Sla})
\begin{equation}
\left|
\begin{array}{ccc}
\tilde{\textbf{V}}_{1}^{*}(\textbf{k})\cdot \textbf{V}_{1}(\textbf{k}) & \cdots & \tilde{\textbf{V}}_{1}^{*}(\textbf{k})
\cdot \textbf{V}_{s_n}(\textbf{k}) \\ 
\vdots & \ddots & \vdots \\ 
\tilde{\textbf{V}}_{s_n}^{*}(\textbf{k})\cdot \textbf{V}_{1}(\textbf{k}) & \cdots & \tilde{\textbf{V}}_{s_n}^{*}(\textbf{k})
\cdot \textbf{V}_{s_n}(\textbf{k})
\end{array} 
\right|.
\end{equation}
The above formula will be used in the evaluation of fidelity for 
the models that we consider.


\section{Results}

In this section we study how the fidelity susceptibility behaves for the mosaic models $ (4-8-8) $ 
and $ (3-12-12) $, 
for the model (\ref{HP}), and for the three-dimensional Kitaev model (\ref{eq:H_3d}).

\subsection{Mosaic models}
Let us first focus on the $(4-8-8)$ mosaic model \cite{mosaic}. 
Choosing a four-site unit cell, a Fourier transform of the Hamiltonian gives 

\begin{equation}
H = \frac{1}{2} \sum_{\textbf{k}} \boldsymbol{\Psi}^{\dagger}(\textbf{k}) \mathcal{H}(\textbf{k}) \boldsymbol{\Psi}(\textbf{k}),
\end{equation}

where 

\begin{equation}
\mathcal{H}(\textbf{k})=\left(\begin{array}{c c}
J_x\sigma^y & -iJ_y\sigma^x + iJ_z\alpha \\
iJ_y\sigma^x - iJ_z\alpha^{\dagger} & J_x\sigma^y
\end{array}\right),
\end{equation}

\noindent $\Psi^{\dagger}_{\mathbf{k}}=(a^\dagger_{\mathbf{k},1},a^\dagger_{\mathbf{k},2},a^\dagger_{\mathbf{k},3},a^\dagger_{\mathbf{k},4})$ 
is the Fourier transform of the Majorana operators $\{\Psi_i\}_{i=1}^4$, $\alpha = \mathrm{diag}[\exp{(-
ik_2)},-\exp{(ik_1)}]$, $k_1 = \mathbf{k}\cdot\mathbf{n}_1$, $k_2 = \mathbf{k}\cdot\mathbf{n}_2$, $\mathbf{n}_1=(1, 
0)$ and $\mathbf{n}_2=(0, 1)$. The system size is $N = 2L^2$.
This model presents a quantum phase transition between two gapped phases with abelian anyons when $J_z^2 = J_x^2 + J_y^2$. 
The phases are algebraically distinct, though they can be related by rotational symmetry.

For the $(3-12-12)$ mosaic model \cite{YaKi, DuScViZa}, 
a Fourier transform with a six-unit cell renders the Hamiltonian in the form 

\begin{eqnarray}\label{Kivelson}
H &=& 
\frac{i}{2}(-a^\dagger_{\mathbf{k},1}a_{\mathbf{k},2} - a^\dagger_{\mathbf{k},1}a_{\mathbf{k},3} +  
Je^{-i\mathbf{k}\cdot\mathbf{n_1}}a^\dagger_{\mathbf{k},1}a_{\mathbf{k},2} \nonumber \\ 
&+& a^\dagger_{\mathbf{k},2}a_{\mathbf{k},3}  - Je^{i\mathbf{k}\cdot\mathbf{n_2}}a^\dagger_{\mathbf{k},1}a_{\mathbf{k},2} - Ja^\dagger_{\mathbf{k},3}a_{\mathbf{k},6} \nonumber \\
 & -& a^\dagger_{\mathbf{k},4}a_{\mathbf{k},5} - a^\dagger_{\mathbf{k},4}a_{\mathbf{k},6} - a^\dagger_{\mathbf{k},5}a_{\mathbf{k},6}) + \mathrm{hc}, 
\end{eqnarray}
\noindent with $\mathbf{n}_1=(1/2, \sqrt{3}/2)$ and $\mathbf{n}_2=(1/2, -\sqrt{3}/2)$. The single particle Hamiltonian $A(\mathbf{k})$ is $6 \times 6$, 
and the system size is $N = 6L^2$.
This system has a quantum phase transition at $J = \sqrt{3}$ from a topological phase with abelian anyons to one with nonabelian anyons, 
spontaneously breaking time reversal symmetry, without the need of an external magnetic field.

\begin{figure}[htp]
\centering
\includegraphics[angle=-90,scale=0.345]{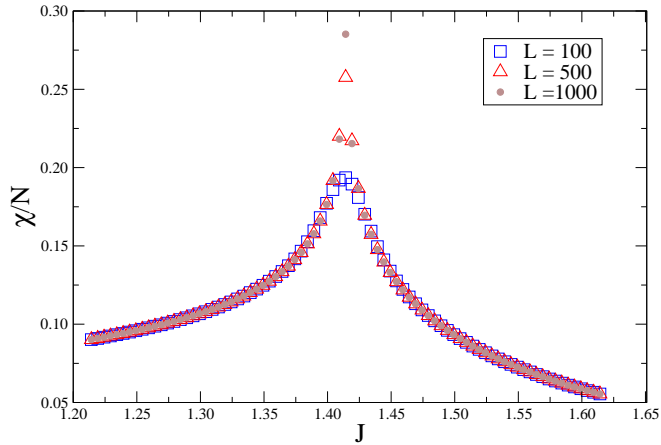}
\caption{(Color online). Fidelity susceptibility $\chi$ per site for the $(4-8-8)$ mosaic model around the quantum critical point $J_z=\sqrt{2}$, 
separating two topological phases ($J_x=J_y=1$). $\chi$ shows extensive behavior off criticality, while it is superextensive at the critical point.}\label{fig:fig3}
\end{figure}

The numerical results for the fidelity susceptibility  $\chi = \lim_{\delta\lambda\rightarrow 0}-2\ln F/\delta\lambda^2$ for the $(4-8-8)$ mosaic model are shown 
in Fig. \ref{fig:fig3} for different system sizes, taking $J_x=J_y=1,J_z\in(1.21, 1.61)$. We see that $\chi$ is an extensive quantity off criticality, 
while it is superextensive at the critical point $J^c_z = \sqrt{2}$.

\begin{figure}[htp]
\centering
\includegraphics[angle=-90,scale=0.34]{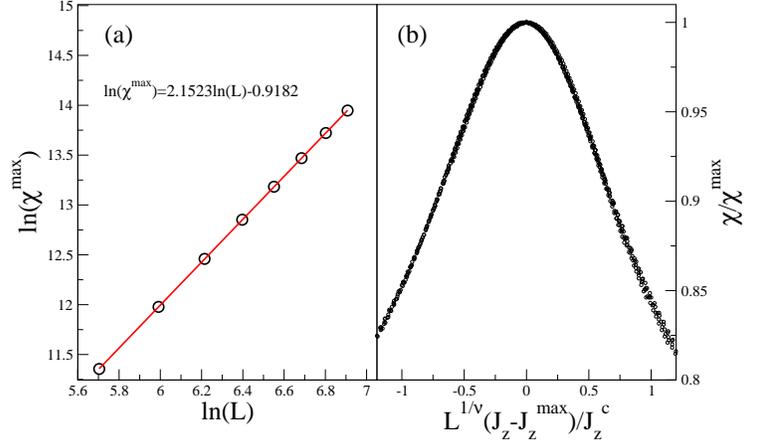}
\caption{(Color online). (a) Scaling analysis of the fidelity susceptibility at criticality for the $(4-8-8)$ mosaic model, for system sizes $L\in [300, 1000]$ in steps of $100$, 
with $\delta J = 10^{-6}$.  We obtain that $\chi^{\mathrm{max}}/N\backsimeq L^\mu$, with $\mu = 0.1523\pm 0.0001$. (b) Data collapse for the fidelity susceptibility around the critical point $J^c_z=\sqrt{2}$. All curves collapse when plotted as a function of the dimensionless quantity $L^{1/\nu}(J_z-J_z^{\mathrm{max}})/J_z^c$, 
with critical exponent $\nu = 1.105 \pm 0.070 $ }\label{fig:fig4}
\end{figure}

We perform finite-size scaling for the susceptibility around the critical point $J_z^c=\sqrt{2}$. For a finite 
sample size we denote the  point at which $\chi$ is maximum as $\chi^{\mathrm{max}}$, evaluated at 
$J_z=J_z^{\mathrm{max}}$. It scales like $\chi^{\mathrm{max}}/N \propto L^\mu$, with $\mu>0$, while in the 
thermodynamic limit $\chi/N \propto 1/|J'-J_z^c|^\alpha$. In Fig. \ref{fig:fig4} (a) we plot 
$\log{\chi^{\mathrm{max}}}$ vs $\log{L}$, for system sizes $L\in [300, 1000]$ in steps of $100$, with $\delta J = 
10^{-6}$. We obtain superextensive scaling given by $\mu = 0.1523\pm 0.0001$.

Furthermore, we perform a data collapse, using the scaling ansatz \cite{BrDaTo}
\begin{equation}
\frac{\chi(J)}{\chi^{\mathrm{max}}}\backsimeq f\Big(\frac{J_z-J_z^{\mathrm{max}}}{\sqrt{2}}L^{1/\nu}\Big).
\end{equation}

To quantify the extent of the collapse, we use a method similar to the one described in 
Ref.\cite{distfss}, obtaining the 
value $\nu = 1.105 \pm 0.07$. 
From there, we conclude that $\alpha = \frac{\mu}{\nu} = 0.138 \pm0.009$. The result for the data collapse is shown in Fig. \ref{fig:fig4} (b).

\begin{figure}[htp]
\centering
\includegraphics[angle=-90,scale=0.35]{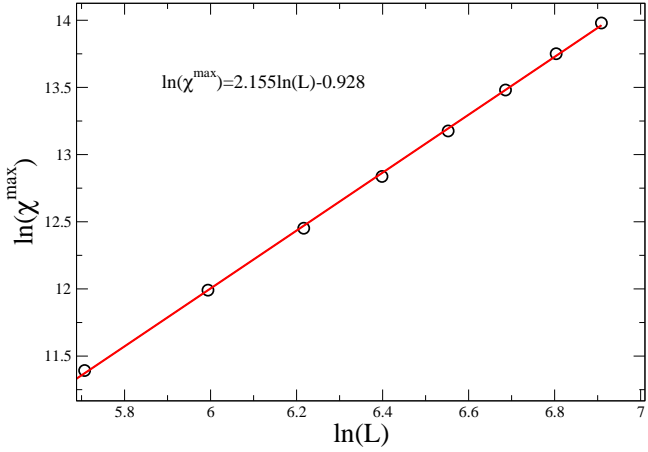}
\caption{(Color online). Scaling analysis of the susceptibility at criticality for the $(3-12-12)$ mosaic model, for $L\in [301, 1001]$ in steps of $100$. 
The fidelity signals the quantum phase transition separating the two distinct topological phases at $J=\sqrt{3}$,
 behaving as $\chi^{\mathrm{max}}/N\backsimeq L^\mu$, with $\mu = 0.155\pm0.009$}\label{fig:fig5}
\end{figure}

A similar analysis has been performed for the $(3-12-12)$ mosaic model \cite{YaKi}, 
confirming that the fidelity is superextensive at criticality, with $\mu=0.155\pm0.009$ 
as seen on Fig. \ref{fig:fig5}.

However, we have not been able to obtain a satisfactory data collapse for the $(3-12-12)$ mosaic model, 
probably due to numerical errors.

\subsection{Kitaev model with 3-spin interaction}
This model can be exactly diagonalized using the techniques we have summarized above. In particular, the Hamiltonian 
in equation (\ref{HP}) 
takes the form $H = \frac{1}{2} \sum_{i,j} \hat{A}_{i,j}c_i c_j$, with $\hat{A}_{ij}=iJ_u\hat{Z}_{ij} + K\sum_k\hat{Z}_{ik}\hat{Z}_{jk}$,  
and for the vortex-free case, the spectral matrix in momentum space is

\begin{equation}\label{HPm}
\mathcal{H}(\mathbf{k})= \left(\begin{array}{c c}
g(\mathbf{k}) & if(\mathbf{k})\\
-if^*(\mathbf{k}) & -g(\mathbf{k})
\end{array}\right),
\end{equation}

\noindent where $f(\mathbf{k}) = J_z + J_xe^{i\mathbf{k}\cdot\mathbf{n}_1}+ J_2e^{i\mathbf{k}\cdot\mathbf{n}_2}$, 
$g(\mathbf{k}) = 2K (\sin{\mathbf{k}\cdot\mathbf{n}_1}-\sin{\mathbf{k}\cdot\mathbf{n}_2} + 
\sin{\mathbf{k}\cdot(\mathbf{n}_2 - \mathbf{n}_1}))$, with $\mathbf{n}_1=(1,0)$ and $\mathbf{n}_2=(0,1)$. The system 
size is $N = 2L^2$.

The inclusion of an external magnetic field gives rise to a transition between phases 
with abelian and nonabelian anyon excitations, contained in the region 
$|J_z|\le |J_x| + |J_y|, \; |J_y|\le |J_z| + |J_x|, \; |J_x|\le |J_y|+|J_z|$ of the parameter space.

\begin{figure}[htp]
\centering
\includegraphics[angle=-90,scale=0.35]{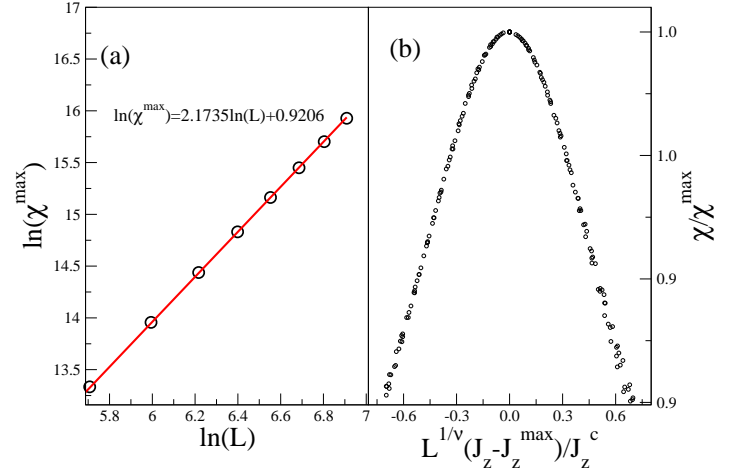}
\caption{(Color online). (a) Scaling analysis of the susceptibility for the Kitaev model with a weak Zeeman magnetic field, at the critical point $J_z=1/2$. 
The system size range is $L\in [301, 1001]$ in steps of $100$, and the coupling $K = 1/15$. The data implies that $\chi^{\mathrm{max}}/N \propto L^\mu$, 
with $\mu = 0.1735\pm 0.0001$ at criticality. (b) Data collapse of the fidelity susceptibility $\chi$ around the critical point $J_z^c=1/2$ 
($J_x=J_y=0.5(1-J_z)$) and $K = 1/15$. We  obtain $\nu=1.10 \pm 0.05$ for the critical exponent of the correlation length.}\label{fig:fig6}
\end{figure}

The numerical results for $\chi^{\mathrm{max}}/N \propto L^\mu$ 
for the honeycomb model with external magnetic field are shown in Fig. \ref{fig:fig6} (a), 
where we obtain $\mu = 0.1735\pm 0.0001$, signaling superextensive scaling at criticality. The corresponding data 
collapse is shown in Fig. \ref{fig:fig6} (b), 
with $\nu = 1.10\pm0.05$ for the critical exponent of the correlation length, and $\alpha =\mu/\nu=0.158\pm0.007$.

For the models we have considered so far we see 
that the fidelity susceptibility is able to detect the phase transition between 
distinct topological phases, besides providing useful information about the critical point itself (i.e. the exponent $\nu$ of the correlation length).

\subsection{3 dimensional model}
The Hamiltonian in equation (\ref{eq:H_3d}) has lines of zeros in momentum space with a vanishing gap. 
The addition of
next-nearest-neighbour interactions that preserve time reversal symmetry of the form 
\begin{equation}\label{eq:Hpert_z}
H_{\rm{nnn}}^{z}=i \sum_{\rm{jk}}t^{z} (\lambda_{j}^{4} \lambda_{k}^{4}-\lambda_{j}^{5}\lambda_{k}^{5})
\end{equation}

\begin{equation}\label{eq:Hpert_x}
H_{\rm{nnn}}^{x}=i \sum_{\rm{jk}}t^{x} (\lambda_{j}^{4} \lambda_{k}^{5}+\lambda_{j}^{5}\lambda_{k}^{4})
\end{equation}
removes the degeneracy \cite{Ryu}. 
In momentum space the Hamiltonian (\ref{eq:H_3d_Mayora}) with the above perturbations 
can be written as
\begin{equation}\label{eq:3d_bz}
H=i \sum_{\rm{k}} (a_{-k}^{4},a_{-k}^{5},b_{-k}^{4},b_{-k}^{5}) \mathcal{H}(k) (a_{k}^{4},a_{k}^{5},b_{k}^{4},b_{k}^{5}),
\end{equation}
with $ \mathcal{H}(k) $ given by 
\begin{equation}\label{eq:single_bz}
\mathcal{H}(\textbf{k})= \left( \begin{array}{cc}
\Theta (\textbf{k}) & i \Phi (\textbf{k}) \\ 
-i \Phi^*(\textbf{k}) & -\Theta (\textbf{k})
\end{array} \right).
\end{equation}
Defining the three component vectors $ s_0=1/4(-1,1,-1) $, $ s_1=1/4(1,1,1) $, $ s_2=1/4(-1,-1,1) $ and
$ s_3=1/4(1,-1,-1) $ (see Fig.\ref{fig:fig2}), the blocks of the matrix (\ref{eq:single_bz}) are given by
\begin{equation}\label{eq:phi}
\Phi(\textbf{k})=\sum_{u=0}^{3} \sigma^0 J_u e^{i \textbf{k}\cdot s_u},
\end{equation}
\begin{equation}\label{eq:pert}
\Theta(\textbf{k})=\Theta^{x}(\textbf{k}) \sigma^{x}+\Theta^{z}(\textbf{k}) \sigma^{z},
\end{equation}
\begin{equation}\label{eq:pert_x}
\Theta^{x}(\textbf{k})=t^{x} \left[ \sin{\frac{k_x -k_y}{2}}+\sin{\frac{k_y -k_z}{2}}+\sin{\frac{k_z -k_x}{2}} \right],
\end{equation}
\begin{equation}\label{eq:pert_z}
\Theta^{z}(\textbf{k})=t^{z} \sin{\frac{k_y +k_z}{2}},
\end{equation}
with $\sigma^0$ the two-dimensional identity matrix and $\sigma^x,z$ the usual Pauli matrices.

When all couplings $ J $ are equal, there are still three degenerate points in the 
Brillouin zone at $ (2 \pi,0,0),(0,2 \pi,0) $ and $ (0,0,2 \pi) $. These degeneracies can be made 
massive by adding a small anisotropy in one of the couplings $ J $. Following \cite{Ryu} we set $ t^x=1 $, 
$ J_{2,3,4}=2 $ and $ J_1=2+\delta J_1 $. The free parameters of the model are then $ t^z $ and $ \delta J_1 $. 
The phase diagram presents three topologically distinct phases, distinguished by the winding number. 
For positive $ \delta J_1 $ and positive $ t_z $ the winding number is $ +1 $; for positive $ \delta J_1 $ and 	
negative $ t_z $ the winding number is $ -1 $, and it is zero in the negative $ \delta J_1 $ region \cite{Ryu}. 
There are two topological phase transitions at the two critical lines $ \delta J_1=0 $ and $ t_z=0 $.

\begin{figure}[htp]
\centering
\includegraphics[angle=-90,scale=0.35]{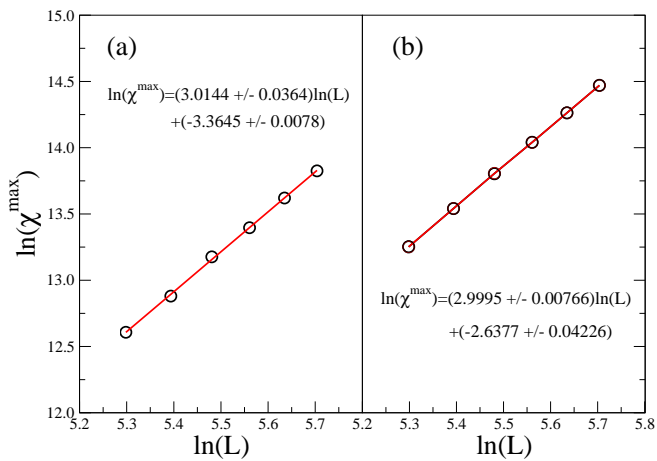}
\caption{ (a) Log-log plot of the fidelity susceptibility at the critical line for $ \delta J_1=1 $, and linear system size $L\in [200, 300]$ in steps of $20$. (b) Log-log plot of the fidelity susceptibility at the critical line for $ t_z=0.1 $, and linear system size $L\in [200, 300]$ in steps of $20$ }
\label{fig:fig7}
\end{figure}

We have studied this model using the fidelity approach and in the following 
we present the results for the scaling of the fidelity susceptibility around the 
two phase transitions of the model. In Fig.\ref{fig:fig7} (a) we show the log-log plot of the fidelity susceptibility 
at the $ t_z=0 $ critical line, while Fig.\ref{fig:fig7} (b) presents analogous results for the transition 
across the $ \delta J_1=0 $ line. In both cases the 
fidelity susceptibility has a peak at the critical point. The scaling analysis suggests that the 
fidelity susceptibility is extensive within the numerical precision that we could obtain 
from a first set of data. However, the numerical accuracy of this result cannot 
exclude a weak super-extensive scaling at the critical point. 
A more accurate analysis of this model will be presented in a future work.

\section{Conclusions}
The fidelity approach proves to be a useful tool to detect and characterize quantum phase transitions which take 
place between distinct topological phases. We have applied it to the family of mosaic models which presents a 
transition between topological phases with abelian and nonabelian anyons, as well as the Kitaev model with an 
external magnetic field, where a transition from an abelian to a nonabelian phase takes place. We were able to 
show superextensive scaling of the fidelity susceptibility at criticality, characterizing its behavior in the 
thermodynamic limit. We also extracted the critical exponent of the correlation length for these models. 
Moreover, we studied the transition between a topological insulator and a topological superconductor in a 
three-dimensional diamond lattice, where the method still detects the boundary between different 
phases. In this case, superextensivity of the fidelity susceptibility turns out to be 
more difficult to be established. 
This is probabily due to the increased dimensionality which weakens the features of the
quantum phase transitions.

We have thus extended the applicability of the fidelity approach to the boundary that separates 
different topological phases and proved its usefulness in this context as well. Since it is 
not always clear how to detect and distinguish different topological phases of matter we believe 
it is important to check the validity of a method based solely on the 
geometrical properties of the 
groundstate, and that does not require the introduction of an order parameter.

\section{Acknowledgements}
We thank K. Shtengel for pointing us reference \cite{YaKi} and H. Yao for helpful discussions. 
Computation for the work described in this paper was supported by the 
University of Southern California Center for High Performance Computing 
and Communications. We acknowledge financial support by the National 
Science Foundation under grant DMR-0804914.

\end{document}